\documentclass[proceedings]{rmaa}
\title{The evolution of disk galaxies}
\author{Vladimir Avila-Reese\altaffilmark{1} and Claudio Firmani\altaffilmark{1,2}}
\altaffiltext{1}{Instituto de Astronom\'{\i}a, Universidad Nacional 
        Aut\'onoma de M\'exico, M\'exico}
\altaffiltext{2}{Osservatorio Astronomico di Brera, Italy}
\fulladdresses{
\item V. Avila-Reese and C. Firmani:
Instituto de Astronom\'{\i}a, UNAM, A.P. 70-268, 04510 M\'exico,
D.F., M\'exico }

\shortauthor{Avila-Reese and Firmani}
\shorttitle{}


\resumen{Estudiamos la evoluci\'on de galaxias de disco utilizando
modelos evolutivos con condiciones iniciales y de frontera asociadas
al escenario de formaci\'on jer\'arquico. Nos concentramos en predecir
la historia de formaci\'on estelar, la evoluci\'on
de tama\~no y brillo superficial y la evoluci\'on de las relaciones
de Tully-Fisher. Presentamos comparaciones con observaciones disponibles.}

\abstract{We study the evolution of disk galaxies using galaxy
evolutionary models with initial and boundary conditions linked to 
the hierarchical formation scenario. We focus our attention on
predictions of the star formation history, size and surface brightness
evolution and the evolution of the $H-$ and $B-$band Tully-Fisher
relations. Comparisons with available observational data are presented.}

\begin{document}

\maketitle

\section{Introduction}

In the last decade, our understanding of galaxy formation and 
evolution grew dramatically thanks to the coming out of 
a theoretical framework for structure formation within the
cosmological context (the inflation-inspired cold dark matter
scenario, CDM), and to the great observational progress,
in particular the imagery and spectroscopy of high redshift
galaxies. Nevertheless, the fundamental questions of galaxy
formation and evolution still await answers. The 
symbiosis of theory and observations is crucial in this
field in order to connect what we ``see''
at different redshifts with the evolution
of a given population of galaxies.

Most of the galaxies observed in the local universe are disk
galaxies. An important question is how the structural, 
dynamical and luminosity properties of this population 
of galaxies evolved and how much they contributed in the past
to global quantities such as the star formation (SF) 
rate and luminosity per unit of volume. 
Were the galaxy disks smaller and their surface
brightness (SB) higher than at present? Which is the SF
history of disk galaxies? Were these galaxies brighter and
bluer in the past? Did the luminosity-velocity (Tully-Fisher)
relation change in the past? A powerful theoretical tool for 
studying these questions related to the evolution of {\it local} 
and global properties of disk galaxies is the combination  
of inductive (backward) galaxy evolutionary models with 
initial and boundary conditions calculated from the hierarchical
formation scenario (Avila-Reese, Firmani, \& Hern\'andez 1998;
Avila-Reese \& Firmani 2000; Firmani \& Avila-Reese 2000).
In this paper, we present some of the evolution predictions of 
these (seminumerical) models and we discuss them in light 
of the available observational data.

\section{Disk galaxy evolutionary models in the hierarchical scenario}

 Disk galaxies are dynamically
fragile objects. This is why, at a first approximation, we 
obviate the disk major mergers in their evolution and 
considering that disks grow inside-out gently with a gas
accretion rate proportional to the hierarchical mass aggregation
rate. A major advantage of our seminumerical models is that 
we follow {\it locally} the overall evolution of {\it individual} 
disks in centrifugal equilibrium, including star formation (SF) and 
feedback in the disk ISM, and luminosity evolution. At each
epoch, the growing disk is characterized by a local infall rate 
of fresh gas, a gas and stellar surface density profile, a 
local SF rate, a color profile, and a rotation curve (including 
the dark matter component). The disks form with the baryon mass fraction
within the growing dark halos. The angular momentum of each collapsing
mass shell is calculated using the spin parameter $\lambda$ obtained
in numerical and analytical works. The local SF is induced
by a gravitational instability (Toomre) criterion and it is 
self-regulated by an energy balance in the ISM along the
vertical disk direction. The efficiency of SF in this model
is almost independent of the mass (luminosity).   

The main physical factors which influence most of the local
and global properties of our model galaxies and their correlations
are the {\bf  mass, mass aggregation history (MAH), and spin parameter 
$\lambda$.} These factors and their statistical distribution are related to 
the cosmological initial conditions. For a quick review of the main 
disk galaxy properties and
correlations predicted at $z=0$ for a typical CDM model, see 
Avila-Reese et al. 2000; the results in detail are presented in 
the papers mentioned above. 

\section{Evolutionary predictions}

\subsection{Star formation rate and luminosity evolution}

The SF rate in our hierarchical models is driven by both the 
{\bf gas accretion rate determined by the MAH} and the {\bf disk
surface density determined by $\lambda$}. In Fig. 1(a) and 1(b)
we show the average hierarchical mass aggregation rate of halos 
of different masses and the corresponding SF rates histories for
$\lambda=0.05$, respectively; less massive galaxies assemble
faster than the massive ones. The correlation between MAH and
SF history is evident. In Fig. 1(b) we also show the SF histories
for $5 \ 10^{11}M_{\odot}$ models with two extreme MAHs 
---early active and extended MAHs (upper and lower dashed 
lines)--- and $\lambda=0.05$. The influence of $\lambda$ on the
SF history is clearly seen in Fig. 1(c) where we plotted 
$5 \ 10^{11}M_{\odot}$ models with $\lambda=0.03$ (upper short-
dashed line) and $\lambda=0.1$ (lower short-dashed line) and
the average MAH in both cases. The upper and lower dot-dashed 
lines correspond to extreme models with an early active MAH 
and $\lambda=0.03$ and with an extended MAH and $\lambda=0.1$, 
respectively.

A strong prediction of our disk galaxy models, due mainly to the 
hierarchical MAHs, is the shape of the SF rate history with
a gentle maximum at $z\sim 1.5-2.5$ for most of the cases,
and a relatively quick fall towards the present by factors
$\sim 2-4$ on the average. From an 
observational study for large disk galaxies ($r_B\ge 4h_{\rm 50}^{-1}$),   
Lilly et al. (1998) have found the SF rate to decrease likely
a factor $2.5-3.5$ between $z\approx 0.7$ and $z\approx 0$ 
(h=0.5). For our corresponding models, since $z\approx 0.7$ the SF 
rate decreased a factor $\sim 2$ on average. However, for 
models with small $\lambda$, i.e., very high SB galaxies, this 
factor is $\sim 3$. It is possible that galaxy samples at 
different redshifts suffer from a significant SB selection effect 
since disk galaxies, as our and other model results show, change 
the SB distribution with $z$ (Bouwens \& Silk 2000). Therefore,
the factor Lilly et al. (1998) find could be biased towards HSB
galaxies in agreement with our galaxy models.

Deep field studies show that the global (cosmic) SF history per unit
of volume from $z\approx 0$ to 0.7 increased by a factor $\sim 6$
and by more than a factor of 10 up to the maximum which is attained
at $z\approx 1.5-2.0$ (e.g., Madau, Pozzetti, \& Dickinson 1998 and the 
references therein). If our models actually describe the evolution
of normal disk galaxies which dominate the global SF rate today,
then the increment of a factor $\sim 10$ detected at $\sim 1.5-2$
could not have been produced only by disk galaxies. Other galaxy
populations had to contribute to the global SF rate in the past
(e.g., Babul \& Fergurson 1996).   

Since the $B-$band luminosity $L_B$ is a tracer of young and intermediate
stellar populations, its evolution should be similar to the SF history.
We find indeed that $L_B$ increases towards the past and more quickly for
the less massive galaxies. At $z=1$, $L_B$ is $\sim 1.5$ times larger
than at $z=0$ for a $L_{\star}$ galaxy. The integral
colors of the galaxy models become bluer towards the past; on average
from $z=0$ to $z=1$, $(B-V)$ decreases by $0.25-0.3$ mag. Less massive
galaxies undergo more color evolution; this is because they attain
the peak in SF rate before than massive galaxies.
\begin{figure}
\vspace{8.7 cm}
\includegraphics{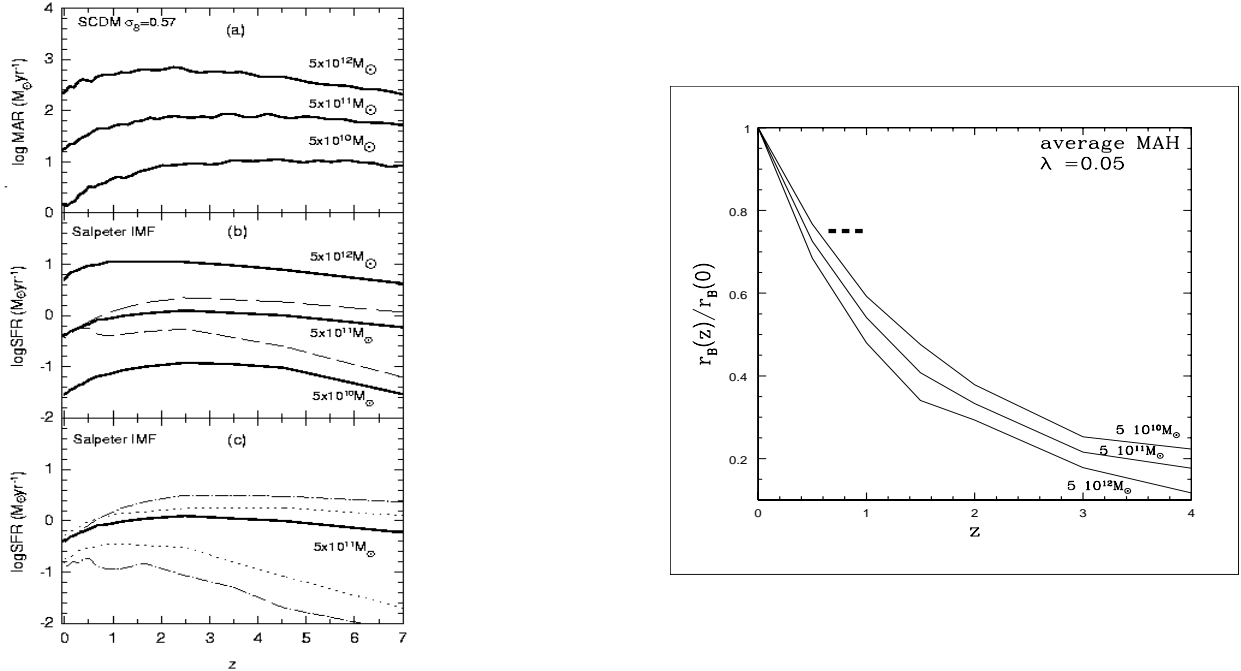}
    \caption{{\it Leftt: }The average mass aggregation rate histories for
3 masses (a), and the corresponding SFHs where $\lambda =0.05$ 
was used (b). The dashed lines are the SFHs of models with
an active early (upper) and extended (lower) MAH. In panel (c), the
dotted lines are for models with $\lambda =0.1$ (upper) and 0.03 (lower);
see text for more details. {\it Right:} Evolution of the $B-$band scale
radius for 3 masses with the average MAH and $\lambda =0.05$. The dashed
line corresponds to an observational estimate for large disk galaxies
by Lilly et al. 1998.} 
\end{figure}

\subsection{Size and surface brightness evolution}

Disk size and SB evolution are natural in the hierarchical 
formation scenario. In the right panel of Fig. 1 we 
show the evolution of the $B-$band scale radius $r_B$ 
for average models of 3 different masses, using the 
flat $\Lambda$CDM ($\Omega _{\Lambda}=$h=0.7) 
cosmology. The size evolution is slightly more
pronounced for more massive galaxies. In this figure we also plot
a rough observational estimate by Lilly et al. (1998). According
to theses authors, for large disk galaxies $r_B$ has grown
not more than $\sim 25\%$ since $z\sim 1$.

Size evolution implies SB evolution. Lilly et al. (1998; dashed line), 
after correcting
for some selection effects, found that the average SB of their 
large disk galaxies sample has grown $\sim 0.5$ mag at $z\approx
0.7$ with respect to $z\approx 0$. Similar results were obtained
by e.g., Forbes et al. (1996), Vogt et al. (1997), Roche et al. (1998).
The last authors conclude that the deep field observations of
disk galaxies can be better explained by luminosity {\it and size}
evolution models. For our hierarchical models, we find slightly 
more pronounced SB evolution than the observations. In a more recent
observational work, Simard et al. (1999) have found that 
galaxies from $z=0.1$ to $z=0.9$ seem to increase their average SB 
by $\sim 1.3$ mag. However, these authors concluded that if the selection
effect due to comparing low luminosity galaxies in nearby
redshift bins to high luminosity galaxies in distant bins is allowed
for, then no discernible evolution remains in the SB of bright disk 
galaxies. Using the same data of Simard et al., Bouwens \& Silk 
(2000) have derived an increase in the SB of $\sim 1.5$ mag. This 
difference is because the last authors have introduced corrections
for SB selection bias that they find to be important due to the strong
evolution in the SB distribution of disk galaxies. 
Certainly, the disk size and SB evolution are important tests for 
the hierarchical formation scenario.

\subsection{Dynamics: evolution of the $H-$ and $B-$band Tully-Fisher 
relations}

The linking of the resolved photometric and spectroscopic
data to dynamical data is crucial for the understanding of the evolution
and physics of a given population of galaxies. In our hierarchical
models, the mass of the galaxy grows faster than the 
maximum rotation velocity $V_m$; this is because the dense inner parts
of the dark halo are not significantly affected by the ``secondary''
mass infall. The dark halos, and, therefore, the disks, obey
a tight relation between mass and maximum circular velocity (see
e.g., Avila-Reese et al. 1998,1999). For our $\Lambda$CDM cosmology,
we find that the disk stellar mass-maximum rotation velocity 
relation, $M_d=AV_m^n$, at $z=0$ has a slope $n\approx 3.4$. This
relation is proportional to the $H-$band Tully-Fisher relation (TFR).
The slope $n$ remains almost constant with $z$. The
coefficient $A$ decreases with $z$, i.e., as we said above,
while the mass significantly decreases towards the past, $V_m$ 
decreases only a little. We find that at $z=1$, $A$ is $\sim 1.6$
times smaller than at $z=0$ or, that is the same, the zero-point
of the $H-$band TFR is $\sim 0.5$ mag higher.

In the $B-$band, the TFR evolves in an opposite way: although the disk
mass ($L_H$ luminosity) decreases with $z$, $L_B$ increases as we 
have seen in $\S 3.1$. This is because SF is more active in the
past. The slope of the $B-$band TFR slightly decreases with $z$
(less massive galaxies evolve more quickly than massive ones;
in particular, $L_B$ increases more with $z$ for the former than
for the latter). Assuming the slope constant with $z$, for our 
$\Lambda$CDM model, we obtain
that the zero-point of the $B-$band TFR at $z=1$ is $\sim 0.7$ mag 
lower than at $z=0$ (taking into account HSB and LSB galaxies),
i.e. $A$ is $\sim 2$ times larger.

From the observational point of view, several efforts were made
in order to acquire internal kinematic (Vogt et al. 1997 and more 
references therein) and galaxy-galaxy lensing (Hudson et al. 
1998) data for disk galaxies at high redshifts. From these
works it is still difficult to draw conclusions regarding the
evolution of the $B-$band TFR. Since the interpretation of the
results is sensitive to the cosmological model assumed, just in order
to evaluate the data from several of these works, we used
the critical model ($q=0.5$) with h=0.5. The results of 
Vogt et al. (1997) show that the zero-point of the $B-$band TFR,
from $z=0$ to the average redshift of the sample $\langle z \rangle=0.54$, 
has changed by $\Delta M_B\approx 0.33$ mag; for the 
Hudson et al. (1998), Simard \& Pritchet (1998) and Rix et al. 
(1997) samples, $\langle z \rangle=0.6,$ 0.35, and 0.25, and 
$\Delta M_B\approx 1.0, -1.5$ and $-1.5$ mag, respectively.
As one sees, these works seem to be at odds with one another. 
The samples used in the last two papers are dominated by small, 
actively star-forming galaxies, while in the two first papers, 
normal spiral galaxies dominate. Thus, we may compare
our model results with those of Vogt et al. and Hudson et al.
papers. For the SCDM cosmology (with $\sigma _8=0.67$ and h=0.5),
$\Delta M_B\approx -0.4$ mag at $z=0.6$ (in fact, the theoretical 
evolution of the TFR is similar for both the SCDM and 
$\Lambda$CDM models). This strongly disagrees with observations in a 
$q=$h=0.5 cosmology. For low density or flat with cosmological
constant models with h$>0.5$, the observational data are in 
better agreement with the models. In particular, in the flat
$\Lambda$CDM cosmology ($\Omega _{\Lambda}=$h=0.7), the data of 
Vogt et al. give only a slightly larger value for $\Delta M_B$ 
than our models (i.e., the observed zero-point at 
$\langle z \rangle=0.54$ is slightly less bright than our hierarchical 
models predict). 

{\bf In conclusion, we find that the hierarchical formation scenario
(for $\Lambda$CDM-like models) offers robust initial and boundary 
conditions for disk galaxy evolution.} However, the evolution of disk
sizes and the zero-point of the $B-$band TFR seems to be too 
exagerated with respect to some observational data. We stress
that analytical approaches should take into account that
the zero-point of the ``structural'' and $B-$band TFRs evolve
in a different way.

\end{document}